\nonstopmode
\documentstyle[12pt,aasms]{article}
\newcommand{\Grad}[1]{\mbox{$\nabla #1$}}
\newcommand{\Div} [1]{\mbox{$\nabla\hspace{-3pt}\cdot\hspace{-2pt}#1$}}
\newcommand{\Curl}[1]{\mbox{$\nabla\hspace{-3pt}\times\hspace{-2pt}#1$}}

\newcommand{\DxDy}[2]{\frac{\partial #1}{\partial #2}}
\def\gapprox{\mathrel{\vcenter{\offinterlineskip \hbox{$>$}
    \kern 0.3ex \hbox{$\sim$}}}}
\def\lapprox{\mathrel{\vcenter{\offinterlineskip \hbox{$<$}
    \kern 0.3ex \hbox{$\sim$}}}}
\begin{document}
\setlength{\baselineskip}{12pt}

\title{The Wardle Instability in Interstellar Shocks: \\
I. Nonlinear Dynamical Evolution}
\author{James M. Stone}
\affil{Department of Astronomy, University of Maryland, College Park MD 20742 \\
jstone@astro.umd.edu}

\begin{abstract}
The nonlinear evolution of unstable C-type shocks in weakly ionized
plasmas is studied by means of time-dependent magnetohydrodynamical
simulations.  This study is limited to shocks in magnetically dominated
plasmas (in which the Alfv\'{e}n speed in the neutrals greatly exceeds
the sound speed), and microphysical processes such as ionization and
recombination are not followed.  Both two-dimensional simulations of
planar perpendicular and oblique C-type shocks, and fully
three-dimensional simulation of a perpendicular shock are presented.

For the cases studied here, the instability results in the formation of
dense sheets of gas elongated in the direction of shock propagation,
and oriented perpendicular to the magnetic field.  The formation of a
weak J-type front is associated with the growth of the instability from
an equilibrium shock structure.
After saturation, the magnetic field structure consists of arches which
bow outwards in the direction of shock propagation, and which are
anchored by the enhanced ion-neutral drag in the dense sheets.  In
analogy to the magnetic buoyancy (Parker) instability,  saturation
occurs when the magnetic tension in the distorted field lines is
balanced by drag in the sheets.  For the magnetically dominated shocks
studied here, the distortions in the magnetic field which produce
saturation are very small.  Nonetheless, the enhancements of the ion
and neutral densities in the sheets is very large, between two and
three orders of magnitude compared to the preshock values.  At these
high densities, recombination processes may be important.  The sheets
evolve slowly in time, so that shocks propagating in a homogeneous
medium may leave behind a network of intersecting filaments and sheets
of dense gas elongated in the direction of shock propagation
and perpendicular to the mean field.
The temperature structure and emission properties of
unstable C-type shocks in the nonlinear regime are presented in a
companion paper (Neufeld \& Stone 1997).

\end{abstract} 

\section{Introduction}

Shock solutions to the equations of magnetohydrodynamics (MHD) in a
weakly ionized plasma have been found to contain a rich variety of
structures (Draine \& McKee 1993).  Currently, such solutions are
classified into three types (Draine 1980; Chernoff 1987). `J-type'
shocks contain one or more discontinuous jumps in the fluid properties,
such as density, temperature, or velocity.  `C-type' shock solutions
consist of a continuous transition in all fluid variables mediated by
ion-neutral drag.  `C$^{*}$-type' shocks, a new class
described by Roberge \& Draine (1990), also consist of a continuous
transition in all variables, however in contrast to C-type shocks the
downstream neutral gas velocity is subsonic over a portion of the shock
structure.  Whether a particular shock is J-type, C-type, or
C$^{*}$-type depends upon the shock speed $v_{s}$, and the properties
of the ambient medium into which it propagates.  Especially important
in this respect is the cooling rate of the gas; shocks in gas in which
the cooling rate is high enough so that the cooling length is small compared to
the ion-neutral coupling length are likely to be C-type.   Weak shocks
with $v_{s} \lapprox 20 - 50$~km~s$^{-1}$ propagating into dense gas
with a low ionization fraction are generally thought to be C-type
(McKee, Chernoff \& Hollenbach 1984).

Observationally, C-type shocks are of great interest for a number of
reasons.  Frictional heating of the neutrals associated with the
ion-neutral drag can excite strong infrared emission lines (Draine,
Roberge, \& Dalgarno 1983; Neufeld \& Melnick 1987; Kaufman \& Neufeld
1996a), or possibly water maser emission (Kaufman \& Neufeld 1996b).
At the same time, strong radiative cooling keeps the gas throughout a
C-type shock cool enough to prevent the dissociation of molecules.  In
fact, rather than destroying molecules, the warm temperatures ($T
\lapprox 4,000$~K) reached by the gas in C-type shocks may instead
drive endothermic reactions, potentially making them sites of rich
interstellar chemistry.  Theoretical models of steady-state planar
C-type shocks provide the best fit to the emission observed towards the
Kleinman-Low nebula in Orion (Draine \& Roberge 1982; Chernoff,
Hollenbach \& McKee 1982; Smith 1991), hot chemistry in molecular
clouds (e.g. Hartquist et al 1990), and have been used to account for
molecular interstellar absorption lines observed in the spectra of some
bright stars (Draine \& Katz 1986).

Interestingly, however, Wardle (1990, hereafter W90; 1991a; b) has
shown that planar C-type shock solutions are unstable to
perturbations in the shape of the shock front above a threshold neutral
Alfv\'{e}nic Mach number $A \equiv v_{s}/v_{A,n}$ (where $v_{A,n}$ is
the Alfv\'{e}n speed in the neutrals) of $A \approx 5$.  The physical
mechanism of the instability bears a close resemblance to the magnetic
buoyancy (Parker) instability in the galactic disk, with the
ion-neutral drag playing the role of gravity.  Small amplitude
corrugations introduced into the shape of the magnetic field lines can
cause the ions to collect in the `valleys' of the field under the
action of the component of drag parallel to the field.  The enhanced
ion density, and therefore drag, in the `valleys' causes the
perturbation to grow, leading to instability.  Through semi-analytic
techniques, Wardle found the maximum growth rate of this instability
increases rapidly with the shock strength: for an $A=10$ shock it is
of order a few times $t_{flow}^{-1}$, where $t_{flow} \equiv (\alpha
\rho_{i})^{-1}$ is the flow time of ions through the shock ($\alpha$ is
the ion-neutral coupling constant, and $\rho_{i}$ the preshock mass
density of ions).  The wavelength of the fastest growing mode is of
order the shock thickness $L_{s} \equiv v_{A,n}t_{flow}$.  For
parameters typical of dense molecular gas, Wardle found $t_{flow}
\approx 10^{3}$~yrs, and $L_{s} \approx 0.01$~pc.  Although Wardle's
calculations were unable to predict the structures produced in the
nonlinear, saturated stage of the instability, it is reasonable to
expect large spatial variations in the ion-neutral drag, and therefore
the heating rate of the gas, in comparison to steady-state planar
solutions.  Such variations may have important effects on the emission
properties and molecular chemistry in the gas.  Without detailed
dynamical calculations of the properties of unstable shocks in the
nonlinear regime, combined with calculations of the resulting emission
properties, the utility of steady-state solutions for planar,
high velocity ($A \gapprox 5$) C-type shocks in modeling the emission
and molecular absorption line observations in dense interstellar gas is
uncertain.

The first calculations of the nonlinear stage of the Wardle instability
have been presented by T\'{o}th (1994, hereafter T94; 1995).  T94
focused on a discussion of suitable numerical algorithms for following
the time-dependent growth of the instability into the nonlinear
regime.  Simulations of three C-type shocks using different parameters
confirmed the linear growth rates predicted by Wardle, and showed
evidence of large amplitude compression of the ion and neutral
densities into structure elongated with the flow.  T\'{o}th (1995) used
numerical simulations to measure growth rates of the Wardle instability
in regions of parameter space not amenable to the techniques used by
W90, e.g. unstable wavevectors out of the plane defined by ${\bf B}$
and the shock normal.  Compression of the flow into dense structures,
which at times detach from the shock front, was again noted.

Recently, the ZEUS code (Stone \& Norman 1992a; 1992b) has been
extended to allow the study of the dynamics of partially ionized
plasmas.  This new version of ZEUS, hereafter referred to as ZEUS-2F,
implements a numerical algorithm which solves the full set of
conservation laws for the coupled ion and neutral fluids, including
time-implicit differencing of the collisional drag terms.  Although
MacLow et al (1995) have described a method for incorporating
the effects of ambipolar diffusion into the ZEUS code, their extension
is applicable only in the strong coupling limit, and assuming
ionization equilibrium holds throughout (so that the ion density simply
scales as a power of the neutral density).  While these assumptions are
useful in a wide variety of circumstances, they preclude the study of
the Wardle instability in C-type shocks, which motivates the extensions
to the ZEUS code described here.  (MacLow \& Smith 1997 have independently
developed similar extensions to the ZEUS code to study C-type shocks.)

In this paper, ZEUS-2F is used to study the nonlinear evolution in
two-dimensions of unstable C-type shocks which propagate in a direction
which is either perpendicular or oblique to the ambient magnetic field
direction.  Moreover, for the first time, a fully three-dimensional
simulation of an unstable C-type shock is presented.  In each case the
evolution is followed far into the nonlinear regime, allowing the
saturation mechanism to be identified.  In all cases studied,
very large density enhancements (by factors of several hundred) in the
postshock gas are produced in the nonlinear regime, with corresponding
per-particle heating rate enhancements of a factor of a few.  In a
companion paper (Neufeld \& Stone 1997, hereafter Paper II), the
effect of these variations on the temperature distribution and
line emission from unstable C-type shocks is examined in detail.

The organization of this paper is as follows.  In the following section,
the numerical algorithms, methods for computing initial C-type shock
structure, and parameters used in the simulations are described.  In \S3
both two- and three-dimensional simulations of unstable C-type shocks
are presented, while conclusions are made in \S4.

\section{Methods}

\subsection{Numerical Algorithms}

In a partially ionized plasma, the ions and neutrals obey separate
systems of conservation laws that are coupled only through a frictional
drag term.  Following Draine (1980), the evolution equations can
be written as
\begin{equation}
 \DxDy{\rho_{i}}{t} + \Div{\rho_{i} {\bf v}_{i}} = 0,
\end{equation}
\begin{equation}
 \DxDy{\rho_{n}}{t} + \Div{\rho_{n} {\bf v}_{n}} = 0,
\end{equation}
\begin{equation}
\rho_{i} \left( \DxDy{{\bf v}_{i}}{t} + {\bf v}_{i} \cdot \Grad{{\bf v}_{i}} 
 \right) = - \Grad{P_{i}} + \frac{1}{4 \pi} (\Curl{\bf B}) \times {\bf B}
 + \alpha \rho_{i} \rho_{n} ({\bf v}_{n} - {\bf v}_{i}) ,
\end{equation}
\begin{equation}
\rho_{n} \left( \DxDy{{\bf v}_{n}}{t} + {\bf v}_{n} \cdot \Grad{{\bf v}_{n}} 
 \right) = - \Grad{P_{n}}
 - \alpha \rho_{i} \rho_{n} ({\bf v}_{n} - {\bf v}_{i}) ,
\end{equation}
\begin{equation}
 \DxDy{\bf B}{t} = \Curl{({\bf v}_{i} \times {\bf B})}.
\end{equation}
Here, each symbol has its usual meaning with the subscript $i$ ($n$)
denoting ion (neutral) variables.  The collisional coupling constant
$\alpha$ is assumed to be a constant;
any dependence of
$\alpha$ upon the drift velocity, ($v_n - v_i$) is neglected.  The strength of
the collisional coupling is conveniently characterized by the
characteristic shock length scale, 
$L_s \equiv V_{A,n}^0/[\alpha \rho_i^0]$, where $V_{A,n}^0$
is the Alfv\'{e}n speed in the preshock neutral fluid and $\rho_i^0$
is the preshock density of the ionized fluid.  A flow timescale may also 
be defined, $t_{flow} \equiv  L_s / V_s = 1/[\alpha \rho_i^0 A]$,
where $A$ is the Alfv\'{e}n Mach number for the neutral fluid.
Note that $\alpha$ appears only as the product $\alpha \rho_{i}$
in [3] and [4].

There are a number of assumptions which have been made in writing
equations [1]-[5].  Firstly, ionization and recombination processes are
neglected, so that there are no source or sink terms in equations
[1]-[2].  This assumption will be valid in the linear regime provided
the growth time of the instability is short compared to ionization and
recombination times.  However, in regions of very high density formed
in the nonlinear stage of the instability, this assumption may break
down.  Including a realistic treatment of ionization and recombination
is difficult due to uncertainties in the dominant microphysical
processes themselves, and because dynamical simulations at a realistic
value of the ion fraction in dense gas (i.e. $\rho_{i}/\rho_{n} \lapprox
10^{-7}$) are very difficult.  The potential effects of recombination
on the saturation of the instability are discussed more fully in \S3.  A
second assumption is the use of an isothermal equation of state for
both the ions and neutrals, so that $P_{i}/\rho_{i} = P_{n}/\rho_{n} =
C_{s}^{2}$, where $C_{s}$ is the isothermal sound speed.  This avoids
requiring a detailed treatment of the complex cooling processes in the
molecular gas, however this assumption limits this study to shocks in
magnetically dominated gases, i.e. $V_{A,n} \gg C_{s}$ in the preshock
gas (in fact, this condition holds in most regions of the ISM), so that
gradients of the thermal pressure are negligible compared to drag and
Lorentz forces.  In Paper II, calculations of the microphysical
molecular cooling processes are performed to compute the equilibrium
temperature of gas in the nonlinear stage of the models presented here:
these calculations show that even in the nonlinear regime thermal pressure
gradients are negligible.  Studies of the dynamics of shocks in weakly
magnetized gases, where $V_{A,n} \leq C_{s}$, will require a proper
treatment of the thermodynamics of the gas.

Numerical methods are required to study the nonlinear dynamics of
unstable C-type shocks: how can such methods be constructed?  Ignoring
for the moment the collisional drag term in equations (3) and (4), it
is clear the neutral evolution equations (2) and (4) are merely the
standard equations of hydrodynamics, while the ion evolution equations
(1), (3), and (5) are the equations of ideal MHD.  Thus, the
hydrodynamical algorithms described in Stone \& Norman (1992a) can be
used to evolve the neutrals, while the MHD algorithms described in
Stone \& Norman (1992b) can be used to evolve the ions, with in both
cases the drag term operator split from the solution of the rest of the
equations.  Because the ion-neutral drag terms potentially are very
stiff, explicit time differencing of these terms are limited by what
may be a very restrictive stability criterion.  Previous authors have
chosen different solutions to this problem: in T94 spatial averaging
was used to provide stabilizing diffusion, while MacLow et al (1995)
used subcycling on the drag terms at a timestep very much smaller than
that dictated by the dynamics.  Here, time implicit differencing is
used to ensure the stiff drag terms are unconditionally stable.
Because the drag term does not involve spatial gradients and is linear
in the velocities, implicit
difference formula are particularly simple in this case.  Let ${\bf
v}_{i}^{n}$ denote the ion velocity at the current time level $n$,
while ${\bf v}_{i}^{n+1}$ denote the value at the advanced time level
$n+1$ (with ${\bf v}_{n}^{n}$ and ${\bf v}_{n}^{n+1}$ denoting similar
quantities for the neutrals).  Then, it is straightforward to show that
\begin{equation}
{\bf v}_{i}^{n+1} = \frac{{\bf v}_{i}^{n} + \alpha \bigtriangleup t 
 ( \rho_{i} {\bf v}_{i}^{n} + \rho_{n} {\bf v}_{n}^{n})}{1 +
 \alpha \bigtriangleup t (\rho_{i} + \rho_{n})},
\end{equation}
\begin{equation}
{\bf v}_{n}^{n+1} = \frac{{\bf v}_{n}^{n} + \alpha \bigtriangleup t
 ( \rho_{i} {\bf v}_{i}^{n} + \rho_{n} {\bf v}_{n}^{n})}{1 +
\alpha \bigtriangleup t (\rho_{i} + \rho_{n})}
\end{equation}
where $\bigtriangleup t$ is the timestep.  Note that due
to operator splitting, the ion and neutral densities are both
held fixed during the updates [6]-[7].
In the limit of a very large timestep, [6] and [7] converge
to the correct equilibrium solution: the final velocities become
the density weighted averages of the initial values. 
Implicit techniques have also been adopted by T\'{o}th (1995).

Numerical algorithms for the solution of equations (1) -- (5) including
the implicit differencing of the drag terms equations (6) -- (7) have been
implemented in the new code ZEUS-2F.  This
code has been subjected to a wide battery of tests.  By setting
$\alpha = 0$, equations (1) -- (5) become decoupled so that
the hydrodynamic test suite of Stone \& Norman (1992a) can be applied
to the conservation laws for the neutrals, while the MHD test suite
of Stone \& Norman (1992b) and Stone et al (1992) can be applied to
the conservation laws for the ions.  With $\alpha \neq 0$, the code
has been tested as part of this work by checking that it
holds and converges to a steady one-dimensional solution for a C-shock
(see \S2.3), and that it reproduces the linear growth rates for
the instability computed by W90 (see \S3.1).  As part
of a separate investigation of the local dynamics of unstable
partially ionized accretion disks, the code also has been shown to reproduce
the analytic growth rates of the Balbus-Hawley instability in a
partially ionized disk (Hawley \& Stone 1997).

\subsection{Initial Equilibrium Shock Structure}

Simulations of the evolution of unstable C-shocks must begin from an
initial equilibrium shock structure.  These are calculated directly from
equations (1)-(5) assuming a steady-state one-dimensional flow.
Let the superscript 0 denote the upstream (preshock) value of a
variable.  Then, in steady-state, the structure of a C-shock
as a function of position $x$ is given by solving
\begin{equation}
\DxDy{v_{ix}}{x} = \frac{- \alpha \rho_{n} v_{ix} (v_{ix} - v_{nx})}{(v_{ix}^{2}
 - B_{y}^{2}/\rho_{i} - C_{s}^{2})},
\end{equation}
\begin{equation}
\DxDy{v_{iy}}{x} = \frac{- \alpha\rho_{n} v_{ix} (v_{iy} - v_{ny}) -
 \frac{B_{x}B_{y}}{\rho_{i}} \DxDy{v_{ix}}{x}}{v_{ix}^{2} - B_{x}^{2}/\rho_{i}},
\end{equation}
\begin{equation}
\DxDy{v_{nx}}{x} = \frac{\alpha \rho_{i} v_{nx} (v_{ix} - v_{nx})}{(v_{ix}^{2}
 - C_{s}^{2})},
\end{equation}
\begin{equation}
\DxDy{v_{ny}}{x} = \frac{\alpha\rho_{i}(v_{iy} - v_{ny})}{v_{nx}},
\end{equation}
where it can be shown that
\begin{equation}
 \rho_{i} = \rho_{i}^{0} v_{ix}^{0} / v_{ix},
\end{equation}
\begin{equation}
 \rho_{n} = \rho_{n}^{0} v_{nx}^{0} / v_{nx},
\end{equation}
\begin{equation}
 B_{y} = \left\{ \begin{array}{ll}
   B_{y}^{0} \rho_{i} / \rho_{i}^{0} & {\rm if}~ B_{x}=0 \\
   B_{x}^{0} v_{iy} / v_{ix}             & {\rm if}~ B_{x} \neq 0.
   \end{array} \right.
\end{equation}
Equations (8) - (14) represent a coupled system of ordinary
differential equations that can be integrated using standard
techniques.  A fourth-order Runge-Kutta algorithm
is used in this work (see also T94).  For a shock propagating in
the $+x$ direction, integration is begun at large $x$ (a few times
$L_{s}$) with all
variables set to their preshock values, and the system (8)-(14) is then
integrated backwards through the shock until all slopes are zero.  The
numerical integration must be started by adding a small value
$\delta \ll 1$ to the preshock neutral velocity $v_{nx}^{0}$; here
$\delta = 10^{-6}$ is used.  Solutions to the equilibrium shock
structure are specified by choosing values for the four dimensionless
parameters: (1) the ion-neutral mass fraction $f = \rho_{i}/\rho_{n}$,
(2) the sonic Mach number $M = v_{nx}^{0} / C_{s}$, (3) the Alfv\'en
Mach number $A = v_{nx}^{0} / V_{A,n}^{0}$, and (4) the angle between
the shock normal and the magnetic field in the upstream gas $\theta =
\arctan (B_{y}^{0}/B_{x}^{0})$.  Values for these parameters used in
this study are discussed in \S2.3.

Once the structure of a steady C-shock is computed using the above
techniques, it is used as the initial conditions for a time-dependent
evolution computed with the ZEUS-2F code.  Typically, a numerical grid
of size $-3L_{s} \leq x \leq 3L_{s}$ and $0 \leq y \leq 3L_{s}$ is
used, with the shock initially located at $x=0$.  The standard
numerical resolution is 128 by 64 grid points in each dimension.  The
three-dimensional simulations use the same $x$ and $y$ domain, with $0
\leq z \leq 3L_{s}$ in the third dimension.  Periodic boundary
conditions are applied in the $y$ (and $z$) directions, with an inflow
boundary (all variables held fixed at the upstream values) at large $x$
and an outflow boundary (zero slope) at small $x$.  The preshock medium
is given velocity $-v_{s}$, so that the shock structure remains
stationary with respect to the grid, i.e. the simulations are carried
out in the shock frame.  Since C-type shocks are sub-Alfv\'{e}nic in
the ions, it is important to ensure the $x$ domain is large enough so that
Alfv\'{e}n waves in the ions reflected off the upstream boundary are damped
before they can affect the shock structure.  As a test, simulations
using a domain of extent  $-10L_{s} \leq x \leq 10L_{s}$ have been
performed; no discernible differences between the structures which
arise on this larger domain with those presented here are found.

If the initial shock structure is not perturbed,
the ZEUS-2F code will hold the solution arbitrarily long with a maximum
relative error in any variable less than 4\% using 32 grid points per
shock scale length $L_{s}$.  (This error is dominated by a small phase
error which causes the shock to move a fraction of a zone with respect to its
initial position.)  Moreover, the errors converge quadratically
with increasing grid resolution as expected for the spatially second-order
algorithms used in the ZEUS-2F code.  The ability of the code to hold a
steady solution, and the convergence of the errors with resolution serve
as both a test of ZEUS-2F, and as a check on the numerical integration of
(8) - (14).

To seed the Wardle instability, linear amplitude perturbations must be added
to the upstream gas.  Both random zone-to-zone perturbations of the neutral
density with maximum amplitude $\delta \rho_{n}^{0} / \rho_{n}^{0} = 10^{-3}$,
and a cosine perturbation with the same amplitude and a wavelength equal
to the transverse size of the grid are used in the
simulations described here.  There is no discernible difference in the
nonlinear evolution of unstable shocks seeded with random versus cosine
perturbations.   Because of the isothermal equation of state
adopted here, these density perturbations will introduce velocity
perturbations of similar amplitude, ensuring all MHD wave modes are
seeded initially.

\subsection{Parameters}

Time-dependent solutions to equations (1)-(5) are specified by
the three dimensionless quantities: $A$, $M$, and $\theta$.  Conditions
in interstellar clouds considerably restrict the allowable range of
$A$ and $M$.  Given typical neutral Alfv\'{e}n speeds $\sim 2 \rm
\, km \, s^{-1}$ in the dense ISM (Heiles et al. 1993), detailed studies 
of steady-state C-type shocks (Draine, Roberge \& Dalgarno 1983;
Kaufman \& Neufeld 1996a) have shown that the maximum shock
velocity for a C-type shock is $\sim 40 \rm \, km \, s^{-1}$,
corresponding to a maximum Alfv\'{e}n Mach number in the neutral fluid
$A \sim 20$).  Faster shocks are of J-type, due to the effects 
molecular dissociation or self-ionization.  Moreover, the linear 
analysis presented by W90 shows that only shocks with $A > 5$
are unstable, so the range of Alfv\'{e}n Mach number of present interest
is $5 \le  A \le 20$.   Typical gas temperatures in the dense ISM are
at least 10~K, corresponding to a sound speed 
$\sim 0.2 \rm \, km \, s^{-1}$, so the relevant range of $M$ is
$M \le 200$.  Accordingly, C-type shocks described
by the parameter set $A=10$ or 20, $M=40$, 100, or 200; and
$\theta = \pi/2$, $\pi/4$, or $\pi/6$, have been studied.

Our fiducial model with $M=100$ and $A=10$ applies to the case of a 
$\sim 20 \rm \, km \, s^{-1}$ shock propagating in gas of sound
speed $\sim 0.2 \rm \, km \, s^{-1}$ and Alfv\'{e}n speed 
$\sim 2 \rm \, km \, s^{-1}$.  Given a preshock density of
$10^5$ particles per $\rm cm^{-3}$, the characteristic length
scale $L_s \equiv V_{A,n}^0/\alpha \rho_i^0 = 
2.68 \times 10^{15} \rm \, cm$ (Kaufman \& Neufeld 1996a; Paper II),
and the flow time is $t_{flow} \equiv L_s/V_s =  430$ yr.
Note that fixing the shock length scale effectively fixes the product
$\alpha \rho_i^0$.  At these densities, the ion fraction $f \sim 10^{-7}$,
however we find once $f < 10^{-3}$, so that the ion inertia is
unimportant, the dynamical evolution is similar.

Any set of physical parameters which are described by the same
dimensionless parameters values will have the same evolution.  Thus,
the results presented here can be trivially scaled to, for example, a
40~km~s$^{-1}$ shock provided the Alfv\'{e}n and sound speeds in the
ambient gas are $V_{A,n} = 4$~km~s$^{-1}$ and $C_{s}=0.4$~km~s$^{-1}$
respectively.  Moreover, the solutions computed here are can also be
scaled to the conditions representative of diffuse HI gas.  For
example, HI gas with $T=20$~K, $n_{H} = 100$~cm$^{-3}$, $B=7.5$~$\mu$G,
and $V_{s} = 20$~km~s$^{-1}$ result in $M=100$ and $A=10$, i.e. the
fiducial model.  

\section{Results}

\subsection{Two-Dimensional Perpendicular Shocks}

As a fiducial model of an unstable perpendicular C-type shock, a simulation
with parameter values $M=100$, $A=10$, and $\theta=\pi/2$
will be discussed.  The simulation is evolved to a time of $9~t_{flow}$
using the standard resolution of $128\times 64$ zones.
Figure 1 plots grayscale images of the ion and neutral densities at
five times during the transition to the nonlinear regime.
At $t=4~t_{flow}$ dense clumps can be seen to develop in the ions, while
the neutrals show large amplitude perturbations in the shock front.
These clumps quickly develop into very thin sheets elongated in
the direction of the flow.  Dense sheets develop in the neutrals at
the same location.  By $t=7~t_{flow}$, the maximum density in these sheets
is a factor of 280 times larger than the initial postshock density in the
ions, and a factor of 120 times larger in the neutrals.  These values are
substantially higher than the compression in the initial steady-state
shock, which is limited to $\sqrt{2} A$.

Figure 2 plots the time development of the energy associated with the
longitudinal component of the magnetic field, normalized by the initial
field energy.  The plot shows the instability follows a remarkably
exponential growth phase from $t=1~t_{flow}$ to $t=4~t_{flow}$.  The
growth rate of the instability, measured from this graph, is 
$2.4$~$t_{flow}^{-1}$.  Interpolation of the growth rates in Table 1 of W90 to
the parameter values of the fiducial model gives a rate of about
4.6$t_{flow}^{-1}$.  The roughly factor of two discrepancy between these
rates is probably due to the use of periodic boundary conditions
in the transverse direction, which restricts the allowed wavelengths.
At later times, the instability saturates, with no
further field amplification occurring.  The saturation amplitude is
small, only 0.1\% of the initial field energy.  This reflects the fact
the upstream gas is magnetically dominated (with $V_{A,n} >C_{s}$), so
that only a small distortion of the field produces large Lorentz forces
and leads to saturation.  It is possible that in weakly magnetized
gases, thermal pressure gradients will play a larger role in determining
the structure in the saturated state, so that the sheets may be substantially
thicker (however, as discussed in \S2, in this case a detailed treatment
of the cooling processes in the gas is required to treat the thermodynamics
accurately).  

An interesting structure evident at late times ($t>5~t_{flow}$) in
Figure 1 is a narrow shell of dense gas parallel to the shock front at
the ends of the sheets.  The shell appears only in the neutrals.  The
sharp jump in neutral (but not ion) density across the shell is
characteristic of a ``J-type" front (Draine \& McKee 1993).  Indeed,
detailed examination of the profiles of the neutral density and
velocity indicate the former increases by a factor of roughly sixteen
in about 5 zones, while the latter decreases by about four.  For an
isothermal equation of state in the neutrals, these transitions are
consistent with a J-type front with a sonic Mach number of about four.
How does the growth of the Wardle instability in
a C-type shock lead to the formation of a J-type front?  As ions pool
in the valleys of the perturbed field lines, the ion-neutral drag is
substantially reduced in the peaks.  Neutrals entering the shock near
the peaks no longer is decelerated completely by drag before
encountering the slow-moving, postshock neutrals that are present in
the initial equilibrium state.  The interaction of the partially
decelerated neutrals with the postshock gas leads to the formation of
two J-type fronts, one which propagates upstream and provides the
final deceleration of the incoming neutrals, and one which propagates
downstream and sweeps up postshock gas.  Between these two fronts lies the
dense shell evident in Figure 1.

The fact that the evolution of unstable C-type shocks can lead to the
formation of J-type fronts is interesting in and of itself.  In fact,
the formation mechanism outlined above seems generic so that it is
likely J-type fronts will be associated with the evolution of all
unstable C-type shocks, at least in magnetically dominated gases.
However, give the small Mach number of the J-type transitions which
result (only 0.04 of the sonic Mach number of the original C-type shock
in the case of the fiducial model), it is unlikely these fronts will
contribute significantly to the total radiative emission from unstable
shocks.  For example, in the fiducial model, a Mach 4 J-type front will
correspond to a propagation speed of only 0.8~km~s$^{-1}$.  Emission
models from the C-type shocks modeled here are described in Paper II.

By $t=7~t_{flow}$, the transient J-type fronts and associated dense
shell are advected off the grid, and a quasi-steady state structure is
established.  Figure 3 plots the structure of shock at $t=7~t_{flow}$,
deep in the nonlinear regime, in more detail.  Figures 3a and 3b are
contour plots of the ion and neutral density.  Figures 3c and 3d show
the instantaneous ion and neutral streamlines.  Figure 3e plots
representative magnetic field lines, and Figure 3f plots the
per-particle heating rate in the neutrals, $H \equiv
\rho_{i} / \mu (v_{i}-v_{n})^{2}$, where $\mu$ is the mean mass per particle.
These plots reveal a variety of
interesting structures.  The contour plots of the ion and neutral
densities reproduce the structure shown in Figure 1.  The ion and
neutral streamlines show, in both cases, the flow is focused towards
the dense sheets beyond their tips.  The focusing of the ions is
stronger than the neutrals, although in both cases the streamlines
remain very smooth.  Note that instantaneous streamlines will show the
correct orientation of the flow, but give no information about speed.
Moreover, no meaning should be assigned to the density of streamlines,
the plotting package assigns lines automatically.  The plot of magnetic
field lines shows both the compression of the field by the shock
(indicated by an increase in the density of field lines from right
to left), and slight
perturbations in the field especially at the tips of the sheets.  It is
surprising, however, how small the distortions of the field
lines are, given the magnitude of the effects in the density (we explore
the force balance in both the neutrals and ions quantitatively below and
in Figure 4).  Finally,
the per particle heating rate in the neutrals shows the highest heating
rate occurs at the tips of the sheet.  Moreover, the maximum value of
the heating rate is nearly two times larger than the rate in the
initial equilibrium state.  This is in indication the temperature in
the neutrals may be substantially different in the nonlinear regime of
an unstable shock in comparison to a planar steady state model (see
Paper II).

An important goal of this study is to identify the saturation mechanism
of the Wardle instability.  Figure 4 plots the terms in the ion
momentum equation along the dotted line shown in Figure 3a at a time of
$t=7~t_{flow}$.  Note this location is chosen to intersect the sheets
at their tips, where from Figure 4e the distortion of the magnetic
field lines is largest.  Figure 4a plots the $x$-components of: (1) the
ion-neutral drag ($\alpha \rho_{i} \rho_{n}[v_{i,x}-v_{n,x}]$, solid
line), (2) the Lorentz force (dotted line), (3) the pressure gradient
in the neutrals (dashed line), and (4) the neutral inertia (or
equivalently, ram pressure) $\rho_{n}
{\bf v}_{n} \cdot \Grad{{\bf v}_{n}}$ (long dashed line).  Figure 4b plots the
$y$-components of the same terms.
From Figure 4a, in the direction of shock
propagation, the drag and Lorentz forces balance in the ions,
whereas the drag and inertia terms balance in the neutrals, except at
the peaks.  Finite grid effects are primarily responsible for the lack
of balance at the peaks, especially for the Lorentz force, since to construct
the plot simple finite differencing was used rather than the actual
force terms computed using the Method of Characteristics algorithm implemented
in the ZEUS-2F code.  From
Figure 4b, in the transverse direction these same terms balance, though
not as well (e.g., in the neutrals, the drag and inertia terms are
out of balance near the peaks).  As discussed below, this lack of balance in the transverse
direction leads to a slow, quasi-static evolution of the sheets in the
transverse direction.  Both ion and neutral pressure gradients are negligible,
even in the $y$-direction where the sheets are very narrow.  Saturation
occurs when magnetic tension forces due to the distortion of field
lines balances the enhanced drag in the ion sheets.  This result
indicates the nonlinear saturation mechanism of the Wardle instability
is directly analogous to the saturation mechanism of the Parker
instability, with the ion-neutral drag playing the role of gravity.  It
is important to determine what mechanism determines the thickness of
the sheets.  From Figure 4b, the ions are compressed primarily by the
pinch associated with the distortion of the field lines.  A resolution
study of the fiducial model (using 16, 32, and 64 grid points
per shock length $L_{s}$)
shows the sheets are no more than 2-4 zones wide at each resolution,
suggesting their transverse structure is not resolved.  Physically,
pressure gradients can be expected to set a minimum thickness to the
sheets provided $C_{s} \gapprox V_{A,n}$ in the sheets.  If the transverse
magnetic field is compressed by a factor $C$ in the shock, this
requires a density enhancement of $\sim C^{2} (M/A)^{2} \sim 100C^{2}$ in the
fiducial model.  If the sheets are to be resolved with $N$ zones in
the transverse direction in the final state, $100C^{2}\times N$
zones in the transverse zones are required initially
(assuming the density enhancement is a
result of transverse motion).  For any reasonable value of $N$, the
resulting numerical resolution required in the initial state greatly
exceeds practical limits, suggesting that more efficient (e.g. adaptive
grid) techniques are required to resolve the internal structure of the
sheets.  Thus, the maximum density compressions reported for the
simulations presented here should be considered as lower bounds.  Note
that despite the fact that it is unlikely the transverse density
structure is resolved in these simulations, quantities such as the
per-particle heating rate, and pinched geometry of the field lines are
resolved (i.e. they extend over many grid zones), and do not change
with resolution.  (For example, with a resolution of $64 \times 32$ zones,
the structure and evolution of the sheets is similar to the standard
resolution, with the maximum heating rate at the tip of the sheets a 
factor of 20\% smaller.)  Finally, the extreme compressions expected in the
sheets in magnetically dominated shocks may increase the effects of
recombination in the sheets.  In fact, the ion density
in, and therefore structure of, the sheets at saturation may be
determined by microphysical effects, thus it will be important to
incorporate them before pursuing higher resolution simulations.

Note from Figure 4b that the net forces in the transverse direction
do not balance completely, indicating the structure observed in
Figure 1 are not a true steady state.  Indeed, when the fiducial
model is evolved to $t=9$~$t_{flow}$, the two sheets migrate towards
one another and finally merge.  At the same time, a new sheet forms
roughly one $L_{s}$ away (roughly the spacing between the sheets
observed in figure 1).  In the lab frame, the density structure
left by the propagation of an unstable C-shock will therefore consist
of a network of intersecting filaments, rather than purely parallel
sheets.  It is probable that the periodic boundary conditions adopted
in the transverse direction affect the spacing of the sheets, by
restricting the range of available wavenumbers.  Calculation of
the exact pattern of the density structures left behind by a propagating
C-type shock will require simulations with much larger transverse dimensions.

To study the effect of varying the sonic Mach number M on the nonlinear
stage of the Wardle instability, simulations of perpendicular C-type
shocks with values of $M=40$ and 200 have been performed.  In all cases,
varying $M$ does not change the structures produced in the nonlinear
regime qualitatively.  Saturation occurs when the ions and neutrals
are compressed into dense sheets elongated with the flow, although since
the growth rate of the instability increases with $v_{s}$, saturation
occurs at earlier times.  For runs using the standard resolution,
the maximum density in the ions and neutrals is within a factor of
two of that observed for the fiducial model.  Moreover, in each case
the maximum heating rate is increased by similar factors (about two to
four) compared to the heating rate in the initial equilibrium
shock structure.

To study the effect of varying the Alfv\'{e}nic Mach number $A$ on the
solutions, a simulation with $A=20$ and $M=200$ 
has been performed.  Once again, the same sheet-like structures are formed.
The maximum densities in the ions and neutrals are within 50\% of the
values reported for the fiducial model, and the maximum heating
rate is a factor of 3.5 times higher than in the initial equilibrium state.
In summary, our limited parameter survey indicates that the nonlinear
saturation mechanism does not depend on the parameters of the shock, and
moreover the maximum compression and increase in heating rate
are within factors of two over a wide range of variation in $M$ and $A$.

\subsection{Two-Dimensional Oblique Shocks}

Figure 5 plots the structure of an unstable oblique C-type shock
far in the nonlinear regime.  The parameters of this shock are
identical to those of the fiducial model, except $\theta = \pi/6$ now.
The simulation is run using the standard numerical resolution of
$128 \times 64$ grid points.
Figures 5a and 5b are contour plots of the neutral and ion densities.
Figures 5c and 5d show instantaneous streamlines in the ions and neutrals
respectively. Figure 5e plots representative magnetic field lines, and
Figure 5f is the per-particle heating rate in the neutrals (compare
Figure 5 to Figure 3).  The plot is shown at $t=3.8$~$t_{flow}$.
Running the simulations to longer times is prohibitive as the
ion density becomes exceedingly small in some regions of the domain,
leading to divergence of the ion Alfv\'{e}n speed.  It is clear
the simulation presented in Figure 5 is well into the nonlinear
regime, and therefore  serves as a useful example of the structure
of unstable oblique shocks.  Nonetheless, it is also clear that implicit,
or semi-implicit, time integration techniques will be required to
study the long term evolution of oblique shocks (Toth 1995).

Qualitatively, the structures formed by the instability in an
oblique shock are similar to those observed in the perpendicular
shock, e.g. Figure 1.  Both the ions and neutrals are compressed
into dense sheets elongated in the direction of the flow.  The sheets
are thicker in the oblique shock, and are slightly curved.  A dense
shell is evident at the base of the sheets in the neutrals, indicative
of the formation of a J-type front associated with the initial growth
of the instability.  The ion and neutral streamlines show convergence
towards the sheets; in fact backflow (i.e. transverse motions opposite
to the far downstream transverse flow direction) is evident in some
locations.  Note the neutral streamlines show a sharp kink at the
location of the dense shell at the base of the sheets, consistent
with the presence of a J-type front at this location.  The magnetic
field lines show distortions which clearly indicate the dense
sheets in the ions and neutrals are located at `valleys', with `peaks'
in between.  The backflow observed in the ion streamlines occurs
where the magnetic field has opposite slope compared to the far downstream
orientation.  Finally, the per-particle heating rate is strongly
peaked at the tips of the sheets, similar to the perpendicular shock
case.

In an oblique shock, the compression ratio in the initial, steady-state
shock is higher than a perpendicular shock of the same Mach number.
In the nonlinear regime, the maximum compression produced by the instability
in an oblique shock is comparable, though somewhat less, than that
reported in \S3.1 for a perpendicular shock.  At $t=3.8$~$t_{flow}$,
the maximum density  in the neutrals is a factor of 126 times higher
than the preshock value, and a factor of 106 higher for the ions.
This leads to a per-particle heating rate which is a factor of 2.3
higher compared to the highest value in the initial shock.  Again,
such variations may have important consequences on the temperature
structure of an unstable oblique shock (see Paper II).  It is possible
that if the shock were evolved to longer times, the maximum compressions
in the ions and neutrals, and maximum heating rate, may increase over the
values reported here.  As in the case of a perpendicular shock, repeating
the simulation at one half the numerical resolution does not lead to
substantive changes in the maximum compression, or maximum heating
rate reported above.

The pattern of distortion in the magnetic field lines evident in Figure 5e
suggest the saturation mechanism of the instability is the same
in oblique shocks as perpendicular shocks, i.e. magnetic tension
balanced by enhanced drag in the sheets (which acts as an effective
gravity that anchors the field lines).  Indeed, plots of each component
of the forces acting on the ions and neutrals in a horizontal slice
through the sheets similar to Figure 4 confirms this expectation:
in the ions the Lorentz force is balanced primarily by drag, while in
the neutrals drag is balanced by inertia (or ram pressure).  For the
isothermal gas studied here, numerical resolution plays a role in
determining the final thickness of the sheets.  However, the level of
agreement between the components is worse in the transverse direction
than for perpendicular shocks, indicating the structures shown in
Figure 5 are not steady.  Indeed, the evolution preceding
$t=3.8$~$t_{flow}$ shown in Figure 5 shows the sheets are in the
process of merging into one larger structure.  Unfortunately, the
difficulty of following the evolution with time explicit techniques
prevents the study of structures which emerge thereafter.  However,
since a similar merger process is observed in perpendicular shocks on
longer timescales, and in this case the merged sheet is qualitatively
unchanged, there is no reason to expect the merger of sheets in oblique
shocks will result in any new behavior.  We also note that in addition
to merging, the sheets drift transverse to the shock front at about the
Alfv\'{e}n speed, as expected (Wardle 1991b).

In order to examine the effect of varying the obliquity angle,
simulations of unstable shocks with $\theta = \pi/4$ and
$\theta = \pi/3$ have also been performed.  Although the maximum compression
in the initial equilibrium state changes with obliquity angle,
no substantive change in the maximum compression, maximum increase
in heating rate, or qualitative structure of the sheets formed
in the nonlinear regime are found for the obliquity angles considered here.

\subsection{A Three-Dimensional Perpendicular Shock}

To study the fully three-dimensional structure of unstable C-type
shocks, the fiducial model for a perpendicular shock (with parameter
values $M=100$, $A=10$, and $\theta=\pi/2$) has
been evolved on a grid of $188\times 64\times 64$ in the domain
$-3L_{s} \leq x \leq 3L_{s}$, $0 \leq y \leq 3L_{s}$, and
$0 \leq z \leq 3L_{s}$.  Because of the computational expense
of three-dimensional simulations, a parameter survey was not performed,
nor have any oblique shock models been computed in three-dimensions.
Instead, this section will describe the results from this single model.

Figure 6a shows an isosurface of the ion density at $t=9.6$~$t_{flow}$,
using the level $\rho_{i} = 0.02$ ($\rho_{i}^{0} = 0.001$
in the dimensionless units used here).  The view is such that
the shock propagates from right (rear) to left (forward), and the magnetic
field is oriented horizontally (see arrows).  A horizontal slice through the three-dimensional
volume corresponds to the two-dimensional simulation described in \S3.1.
Note that as one might expect, the ions have formed
dense sheets perpendicular to the magnetic field direction.  In fact,
the gross morphology of all the dynamical  variables are similar to
the 2D results: the neutral also form vertical sheets at the same locations
as those in the ions, and, in addition, there is a dense shell of neutrals
located at the base of the sheets associated with J-type fronts.  
Moreover, the ion and neutral streamlines in the horizontal plane
show focusing towards the sheets, and the component of the magnetic field
in the horizontal plane are distorted in a similar fashion to that
shown in Figure 3e.  However, note the sheets are not symmetric in the
vertical direction, but the leading edge of the sheet shows large variations
in $x$-position.  Moreover, the sheets are not perfectly planar, but are
slightly folded in the horizontal direction (this is best traced along
the rear edges of the sheets), i.e. the $y$-coordinate of the leading
edge of the sheet varies with $z$.  Both of these patterns are evidence of
vertical modes.

Figure 6b plots an isosurface of the per-volume heating rate at
the same time as the view in Figure 6a, and at the level $H=20$ (the
maximum heating rate in the initial state is $H^{0} = 5.9$).
Unlike the two-dimensional
case, where the heating rate is strongly localized to the tip
of the sheets, in three-dimensions there is significant heating in
small regions throughout the volume of the sheets.  Significant heating
in horizontal strips along the tips of the sheets is evident.

The complex pattern of the heating rate is an indication that
motions in the third dimension cause significant variations in the
values of flow variables compared to the two-dimensional flow field.
Indeed, at $t=9.6$~$t_{flow}$, the maximum value of the neutral
density is a factor of 339 times higher than the preshock neutral
density, which is significantly higher than the maximum value
in two-dimensions.  The difference in the ion density is even more
pronounced: the maximum ion density is 772 times higher than the
preshock value, nearly an order of magnitude higher than
the result in two-dimensions.  These values lead to a dramatic
increase in the maximum heating rate, a factor of 30 times higher
than the steady-state value.  It is likely that the variation in
temperature structure in a 3D shock is even more pronounced than
in 2D (see Paper II).  At the same time, the high compressions in
the postshock gas (nearly three orders of magnitude compared to
the preshock gas in the case of the ions) indicates the need to
incorporate microphysical processes such as recombination to
correctly model the saturated structure.

Finally, in Figure 7 the time development of the magnetic
energy associated with the components of the field perpendicular
to the initial field direction ($B_{x}$ and $B_{z}$) is plotted.
The growth rates of these components is generally lower than that
measured in the 2D simulation (cf. Figure 2).  The time evolution
of both components are nearly identical, although $B_{x}^{2}$ (the
energy density associated with
the field component in the direction of shock propagation) is roughly
a factor of 10 larger throughout.  An exponential growth phase
between $t=5$~$t_{flow}$ and 8~$t_{flow}$ dominates both components;
the measured growth rate during this phase is 1.9~$t_{flow}^{-1}$.
This value is somewhat lower than the 2D result.  Saturation again
occurs at low amplitude, 0.1\% of the initial magnetic energy.

\section{Conclusions}

Time-dependent MHD simulations have been used to study the nonlinear
evolution of the Wardle instability in C-type shocks in both two- and
three-dimensions.  Shocks which propagate either in a direction
perpendicular or oblique to the direction of the magnetic field have
been studied.  The major conclusions are as follows.

(1) The instability results in the formation of dense sheets of
gas elongated in the direction of shock propagation, and oriented
perpendicular to the initial direction of the magnetic field.  Both
the neutral and ion densities are greatly enhanced in these sheets 
compared to their preshock values, by factors of two to three
orders of magnitude in different circumstances.  The sheets merge,
and new ones are formed, as the shock propagates, leading to the
speculation that on large scales, an unstable C-type shock propagating
through a homogeneous ambient medium will produce a network of
intersecting filaments and sheets of dense gas aligned in the direction
of shock propagation and perpendicular
to the mean field direction.

(2)  Saturation of the instability occurs when magnetic tension
in the magnetic field which arches between the sheets is balanced
by the enhanced drag that anchors the field in the sheets.  In
magnetically dominated shocks, the amplitude of the field distortion
which produces saturation is very small.  For the isothermal
equation of state studied here, the final thickness of the sheets
is unresolved.

(3) J-type fronts are formed during the initial growth phase
of the instability in a planar, steady-state initial shock structure,
because of large variations in the ion fraction across
the surface of the shock front. It is not clear whether
this result will be universally true for the growth of the instability
in all initial unstable states.  However, because
large variations of the ion fraction across
the shock seems an inevitable result of the instability,
the formation of J-type fronts may be a generic feature of the instability.

(4) The per-particle heating rate of the neutrals is increased by
more than a factor of two near the tips of the sheets in two-dimensions,
and by a factor of 30 in three-dimensions.  Calculations of
the temperature and line emission properties of the molecular gas in the
nonlinear stage instability (based on the dynamical models presented here)
is examined in Paper II.

It will be important in future work to relax some of the assumptions
adopted here.  In particular, recombination processes have been
neglected, although they may be important in the nonlinear regime due
to the extreme compression of the gas into the sheets.  Moreover, in
order to study shocks in weakly magnetized gas, a realistic treatment
of cooling will be required rather than the isothermal equation of
state used here.  Finally, studies of shocks in non-planar
geometries (e.g. bow shocks) are important.

\vspace*{1in}  I am grateful to David Neufeld and Mark Wardle for stimulating
discussions.  The computations were performed on the Cray systems at
the Pittsburgh Supercomputing Center, and on the CM-5 system at the
National Center for Supercomputing Applications.  This research was supported
by the NSF through grant AST-9528299.

\newpage
\begin{center}
{\bf References}
\end{center}
\setlength{\baselineskip}{18pt}
\setlength{\parindent}{0cm}
Chernoff, D.F. 1987.  ApJ, 312, 143.

Chernoff, D.F., Hollenbach, D.J., \& McKee, C.F. 1982.  ApJ, 259, L97.

Draine, B.T. 1980.  ApJ, 241, 1021

Draine, B.T., \& Katz, N. 1986.  ApJ, 310, 392.

Draine, B.T., \& McKee, C.F. 1993.  ARA\&A, 31, 373.

Draine, B.T., \& Roberge, W.G. 1982.  ApJ, 259, L91.

Draine, B.T., \& Roberge, W.G., \& Dalgarno, A. 1983.  ApJ, 264, 485.

Hartquist, T.W., Flower, D.R., \& Pineau des For\^{e}ts, G. 1990.  In
{\em Molecular Astrophysics}, ed. T.W. Hartquist, Cambridge, CUP, p91.

Hawley, J.F., \& Stone, J.M. 1997.  In preparation.

Heiles, C., Goodman, A.A., McKee, C.F., \& Zweibel, E.G. 1993, in Protostars
and Planets $\phantom{XXXXX}$III, ed. M.\ Matthews \& E.\ Levy (Tucson: Univ.\ of Arizona Press).

Kaufman, M.J., \& Neufeld, D.A. 1996a.  ApJ, 456, 611.

Kaufman, M.J., \& Neufeld, D.A. 1996b.  ApJ, 456, 250.

MacLow, M.-M., Norman, M.L., K\"{o}nigl, A., \& Wardle, M. 1995.  ApJ, 442, 726.

MacLow, M.-M., \& Smith, M. 1997, submitted to ApJ.

McKee, C., Chernoff, D., \& Hollenbach, D. 1984, in Galactic and Extragalactic
Infrared Spectroscopy, ed. by M.F. Kessler and J.P. Phillips (Dordrecht:
Reidel), 103.

Neufeld, D.A., \& Melnick, G.J. 1987.  ApJ, 332, 266.

Neufeld, D.A., \& Stone, J.M. 1997.  In preparation (Paper II).

Roberge, W.G., \& Draine, B.T. 1990.  ApJ, 350, 700.

Smith, M.D. 1991.  MNRAS, 253, 175.

Stone, J.M., \& Norman, M.L. 1992a.  ApJS, 80, 753.

Stone, J.M., \& Norman, M.L. 1992b.  ApJS, 80, 791.

Stone, J.M., Hawley, J.F., Evans, C.E., \& Norman, M.L. 1992.  ApJ 388, 415.

T\'{o}th, G. 1994.  ApJ 425, 171 (T94).

T\'{o}th, G. 1995.  MNRAS, 274, 1002.

Wardle, M. 1990.  MNRAS, 246, 98 (W90).

Wardle, M. 1991a.  MNRAS, 250, 523.

Wardle, M. 1991b.  MNRAS, 251, 119.

\newpage
\begin{center}
{\bf Figure Captions}
\end{center}

Figure 1.  Time development of the Wardle instability in a $M=100$,
$A=10$ perpendicular C-type shock.  The bottom panels are grayscale
images of the neutral density at times marked in units of $t_{flow} =
(\alpha \rho_{i})^{-1}$.  The top panels show the ion density at the
same times.  The maximum density in the ions (neutrals) is a factor of
280 (120) times larger than the preshock values in the initial,
steady-state.  The size of the computational domain is 6~$L_{s} \times$
3~$L_{s}$.  For a 20~km~s$^{-1}$ shock in dense molecular gas,
with $L_{s} = 2.68\times 10^{15}$~cm,
$t_{flow} \approx 430$~yrs.

Figure 2.  Time history of the magnetic energy in the longitudinal
component (parallel to the shock normal) of the field
normalized by the initial field energy for the shock evolution shown in
Figure 1.  Exponential growth of the instability is evident from
$t=1$~$t_{flow}$ to $t=4$~$t_{flow}$, followed by saturation at a low
amplitude.

Figure 3.  Structure of a $M=100$, $A=10$ perpendicular C-type shock in
the nonlinear regime, at $t=7$~$t_{flow}$.  {\em (a)}  Contours of the
ion density between 0.001 and 0.278.  {\em (b)} Contours of the neutral
density between 1.0 and 119.  {\em (c)} Instantaneous ion streamlines.
{\em (d)} Instantaneous neutral streamlines.  {\em (e)} Representative
magnetic field lines.  {\em (f)} Contours of the per-particle heating
rate in the neutrals between 0.0 and 55.6.  The dashed line in panel
{\em (a)} is the location of the vertical slice used to construct
Figure 4.  Note in the dimensionless units quoted here, $\rho_{i}^{0} =
0.001$, $\rho_{n}^{0} = 1.0$, and the maximum $H^{0} = 22.4 $

Figure 4.  Plots of various terms in the ion and neutral momentum
equations as a function of vertical position $y$ along the slice shown
in Figure 3a.  This position passes through the regions of maximum
heating at the tip of the ion and neutral sheets shown in Figure 3f.
Plotted are the ion-neutral drag (solid line), Lorentz force (dotted
line), pressure gradient in the neutrals (short dashed line), and
neutral inertia term $\rho_{n} {\bf v}_{n} \cdot \Grad{{\bf v}_{n}}$ (long
dashed line). {\em (a)} Terms in the $x$-direction (top panel)
{\em (b)} Terms in the $y$-direction (bottom panel).
Saturation in the ions is a
balance between drag and Lorentz forces, while in the neutrals it is a
balance of drag and inertia.

Figure 5.  Structure of a $M=100$, $A=10$ oblique C-type shock with
$\theta=\pi/6$ in the nonlinear regime, at $t=3.8$~$t_{flow}$.  {\em
(a)}  Contours of the ion density between 0.001 and 0.106.  {\em (b)}
Contours of the neutral density between 1.0 and 127.  {\em (c)}
Instantaneous ion streamlines.  {\em (d)} Instantaneous neutral
streamlines.  {\em (e)} Representative magnetic field lines.  {\em (f)}
Contours of the per-particle heating rate in the neutrals between 0.0
and 103.  Note in the dimensionless units quoted here, $\rho_{i}^{0} =
0.001$, $\rho_{n}^{0} = 1.0$, and the maximum per-particle heating rate
in the initial equilibrium state is 32.5.

Figure 6.  {\em (a)} Isosurface  ($\rho_{i}=0.02$) of the ion density at $t=9.6$~$t_{flow}$
in a 3D simulation of a perpendicular C-type shock.  The arrows
show the directions of the flow velocity and magnetic field.  {\em (b)}
Isosurface ($H=20$) of the per-volume heating rate.  The maximum values
of $\rho_{i}$ and $H$ are 339 and 179 respectively.

Figure 7.  Time history of the magnetic energy in the
components of the field perpendicular to the initial field,
normalized by the initial field energy for the 3D shock evolution shown in
Figure 6.  The solid line plots $B_{x}^{2}$, while the dashed
line plots $B_{z}^{2}$.  Exponential growth of the instability is evident from
$t=5$~$t_{flow}$ to $t=8$~$t_{flow}$.

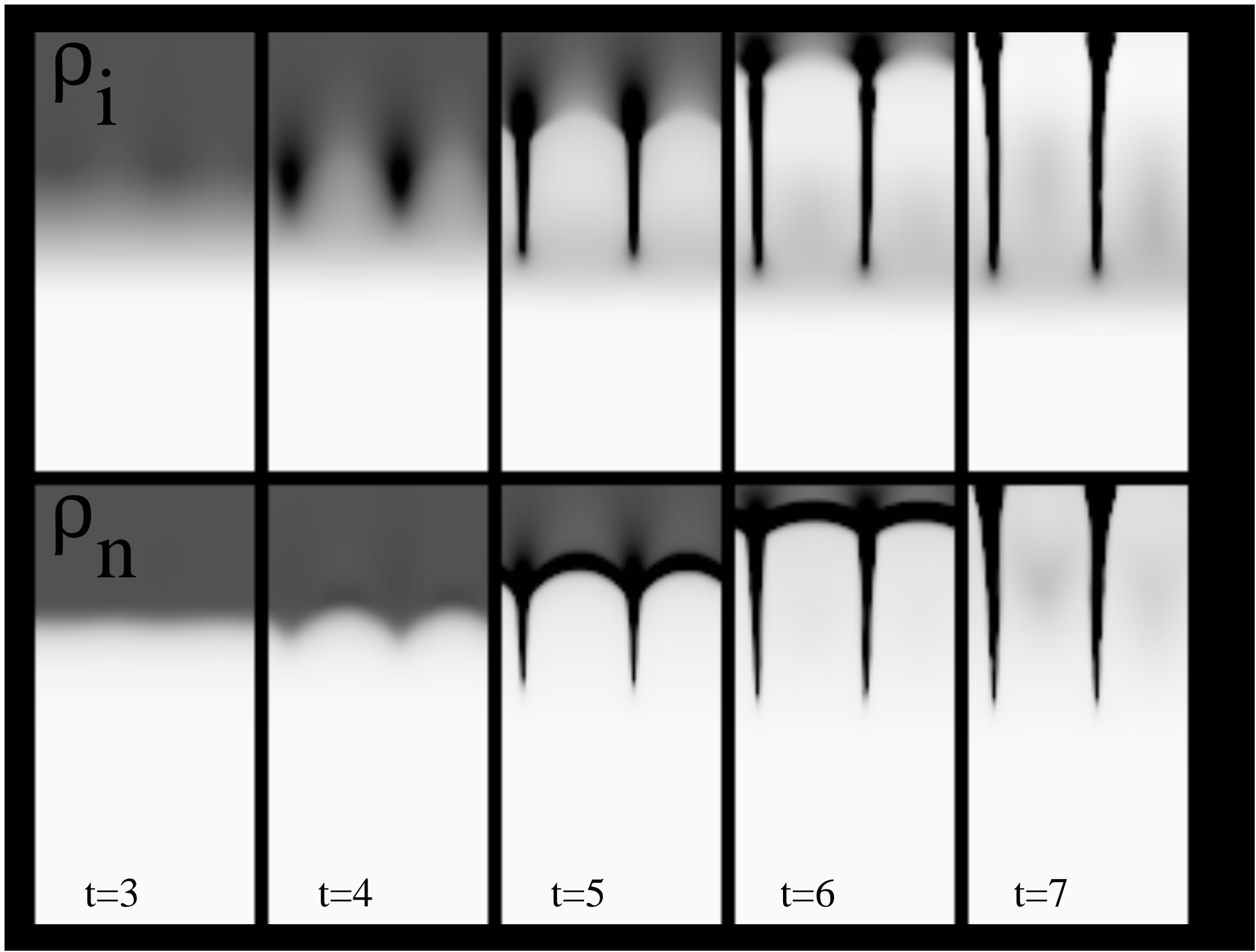
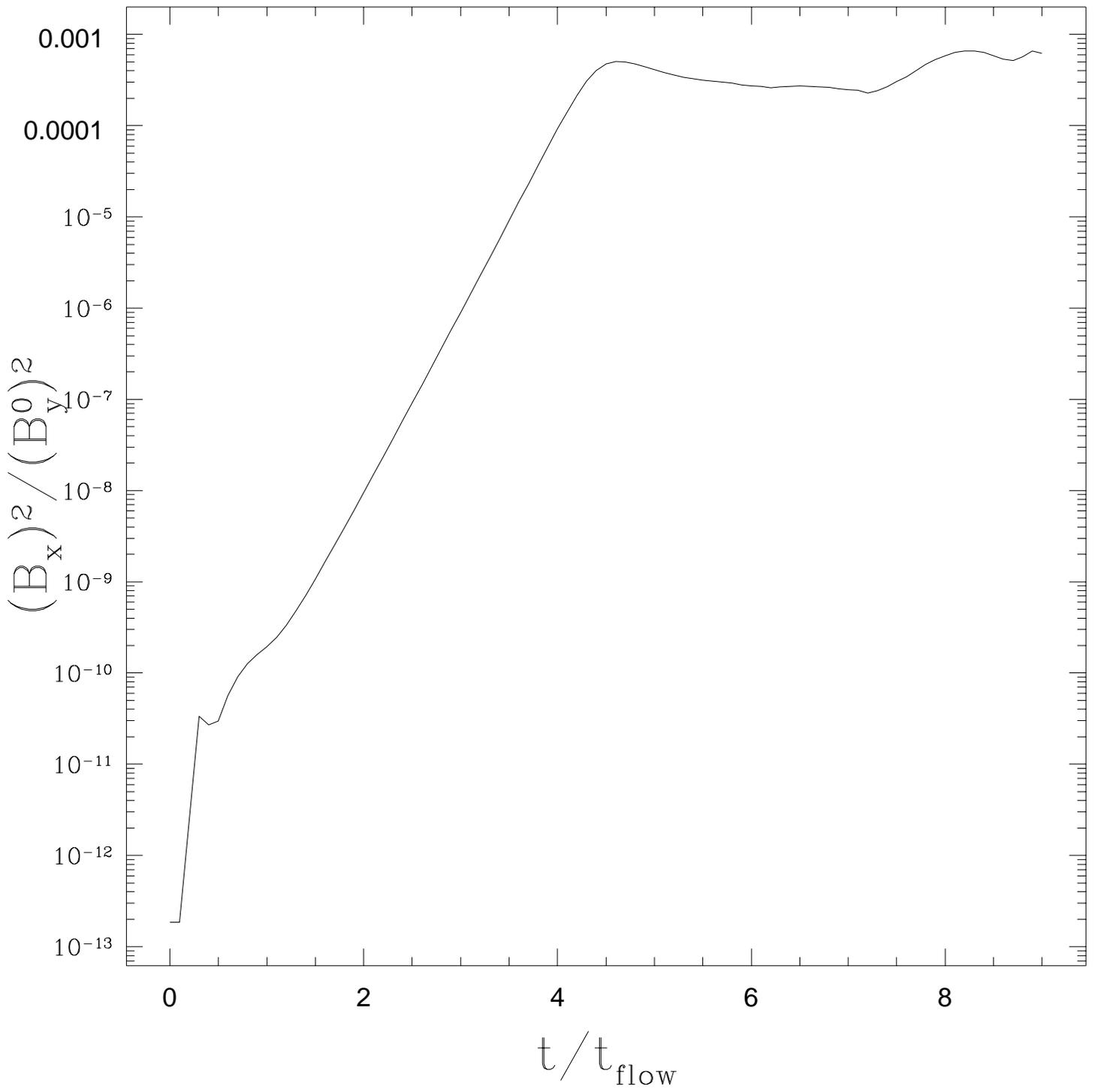
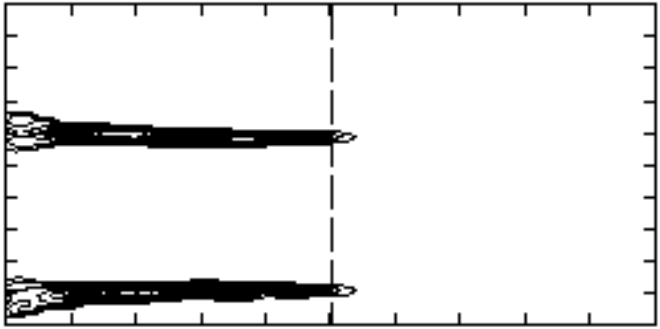
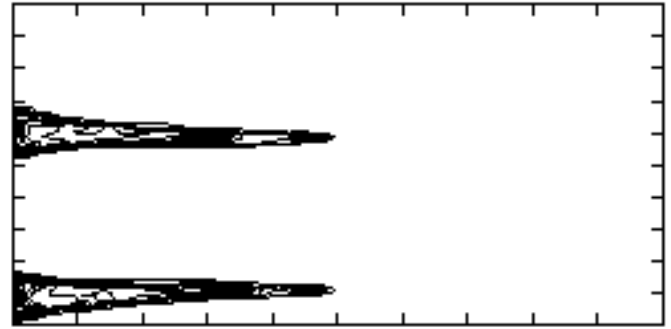
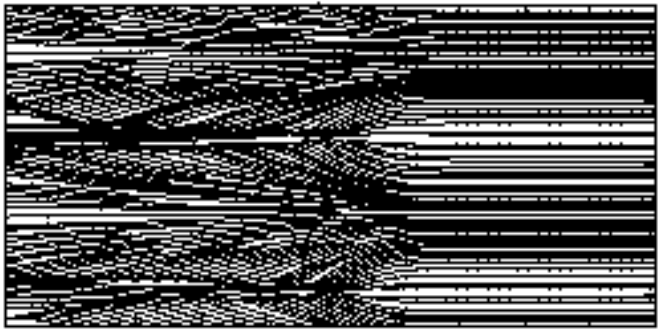
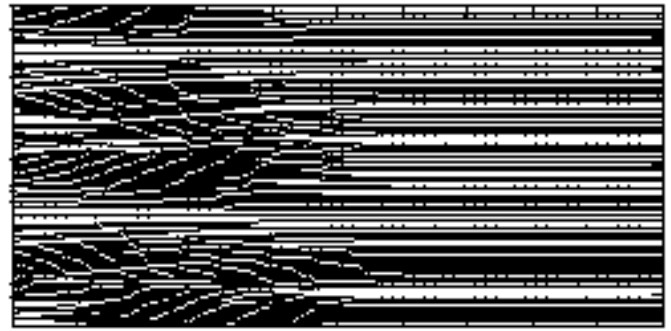
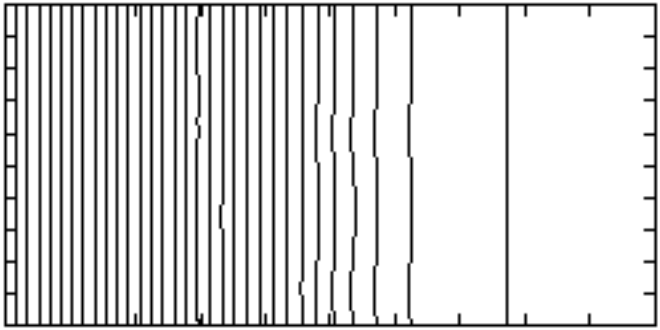
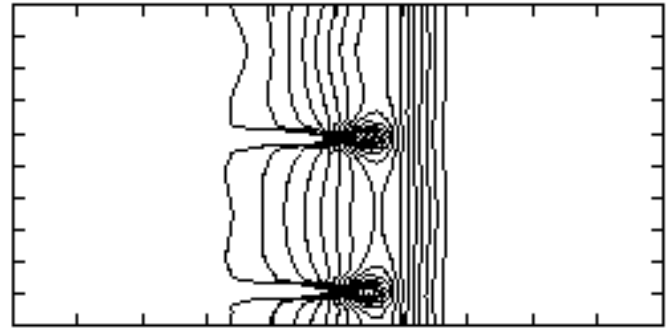
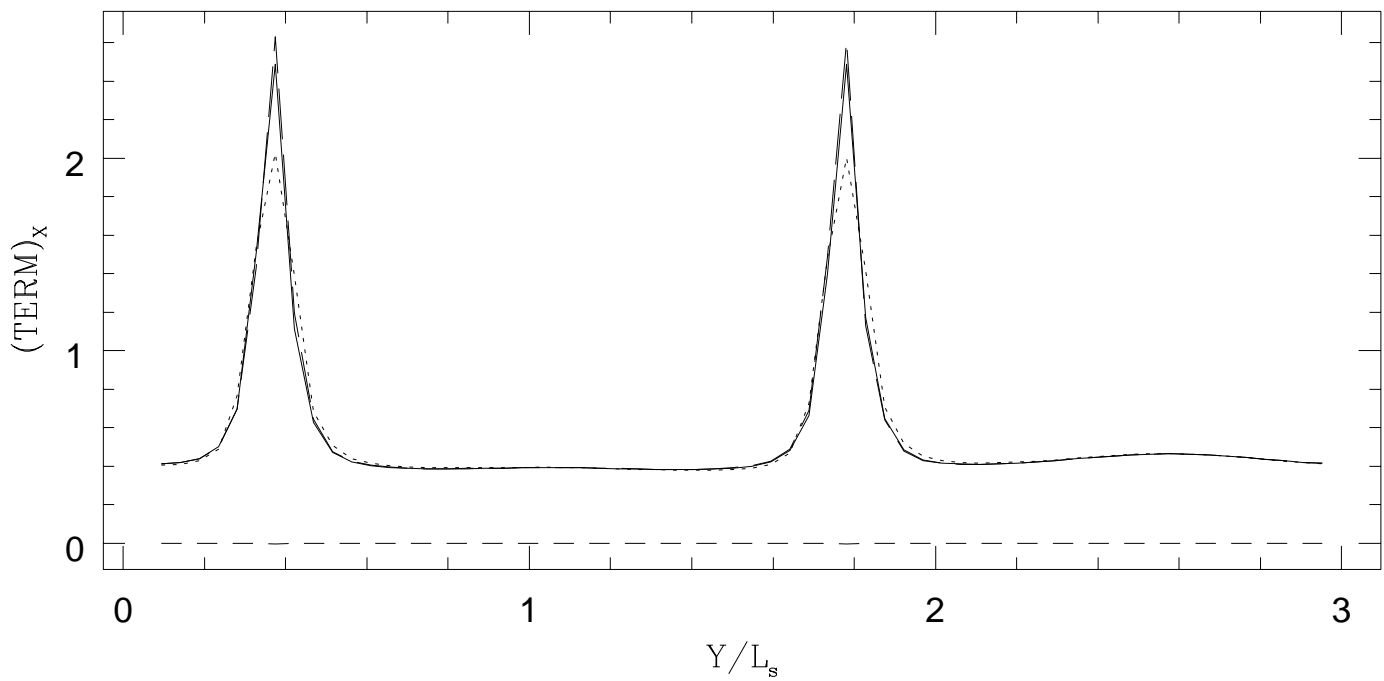
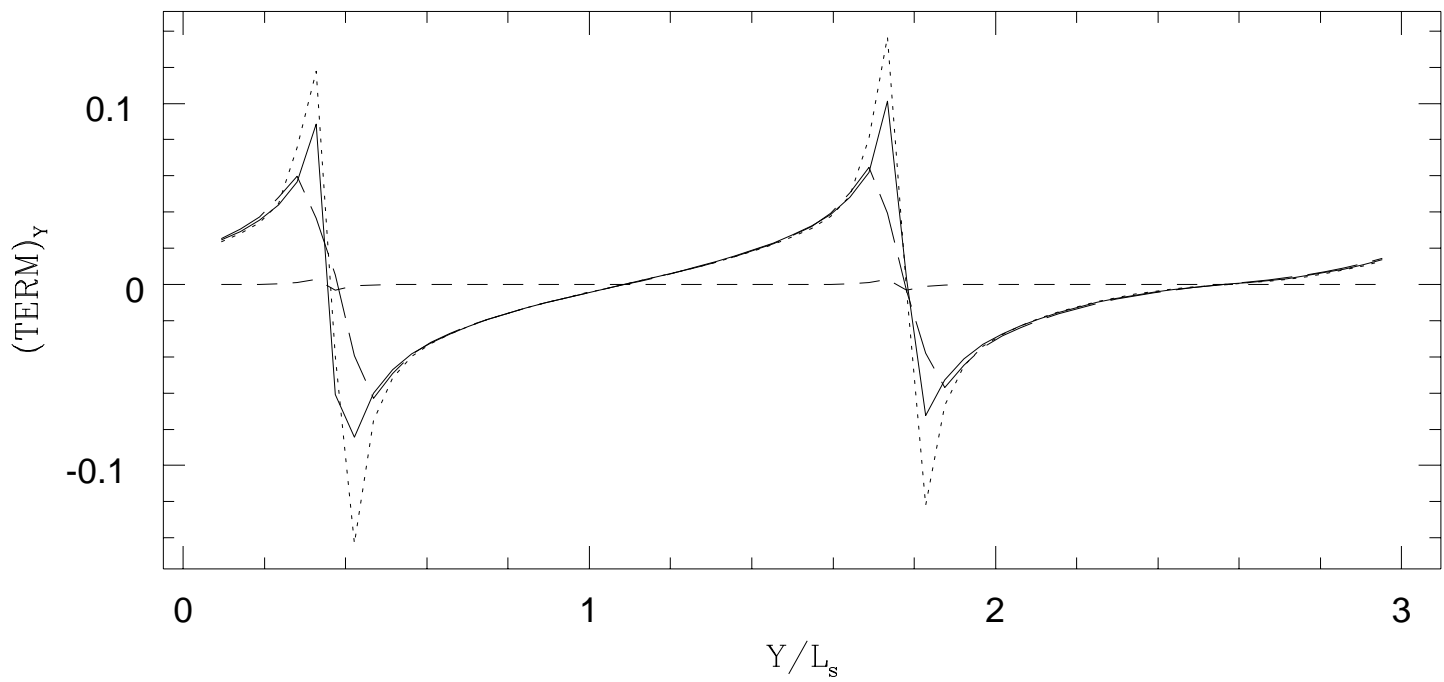
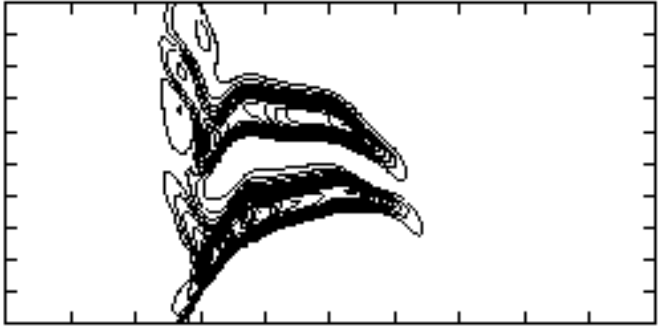
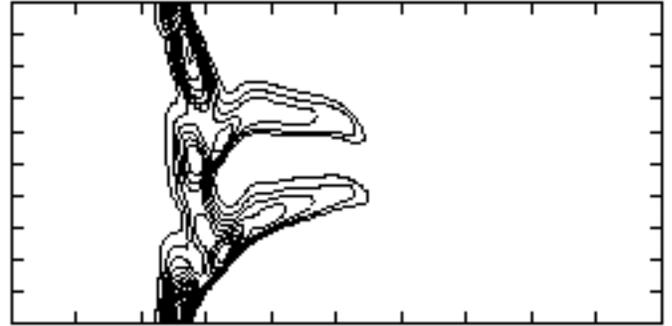
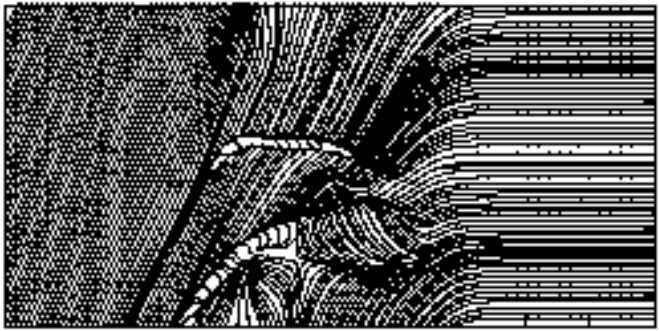
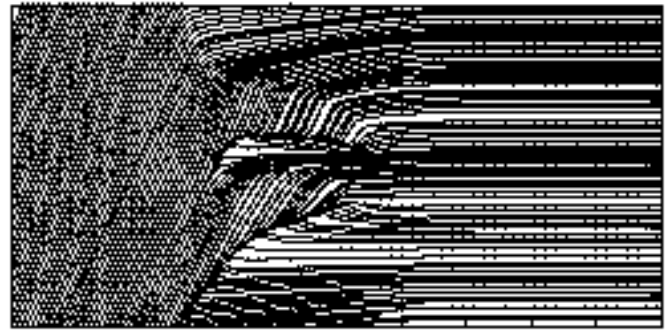
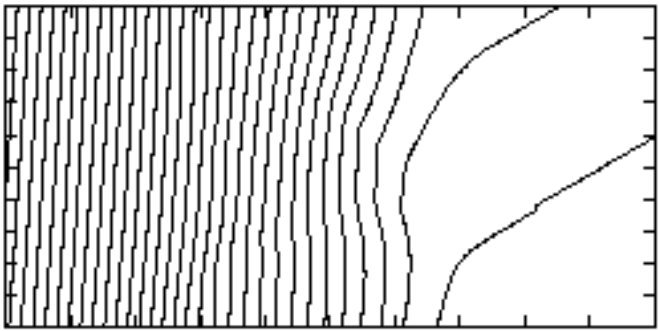
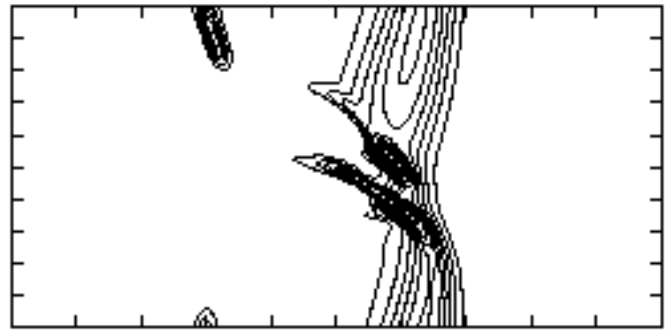
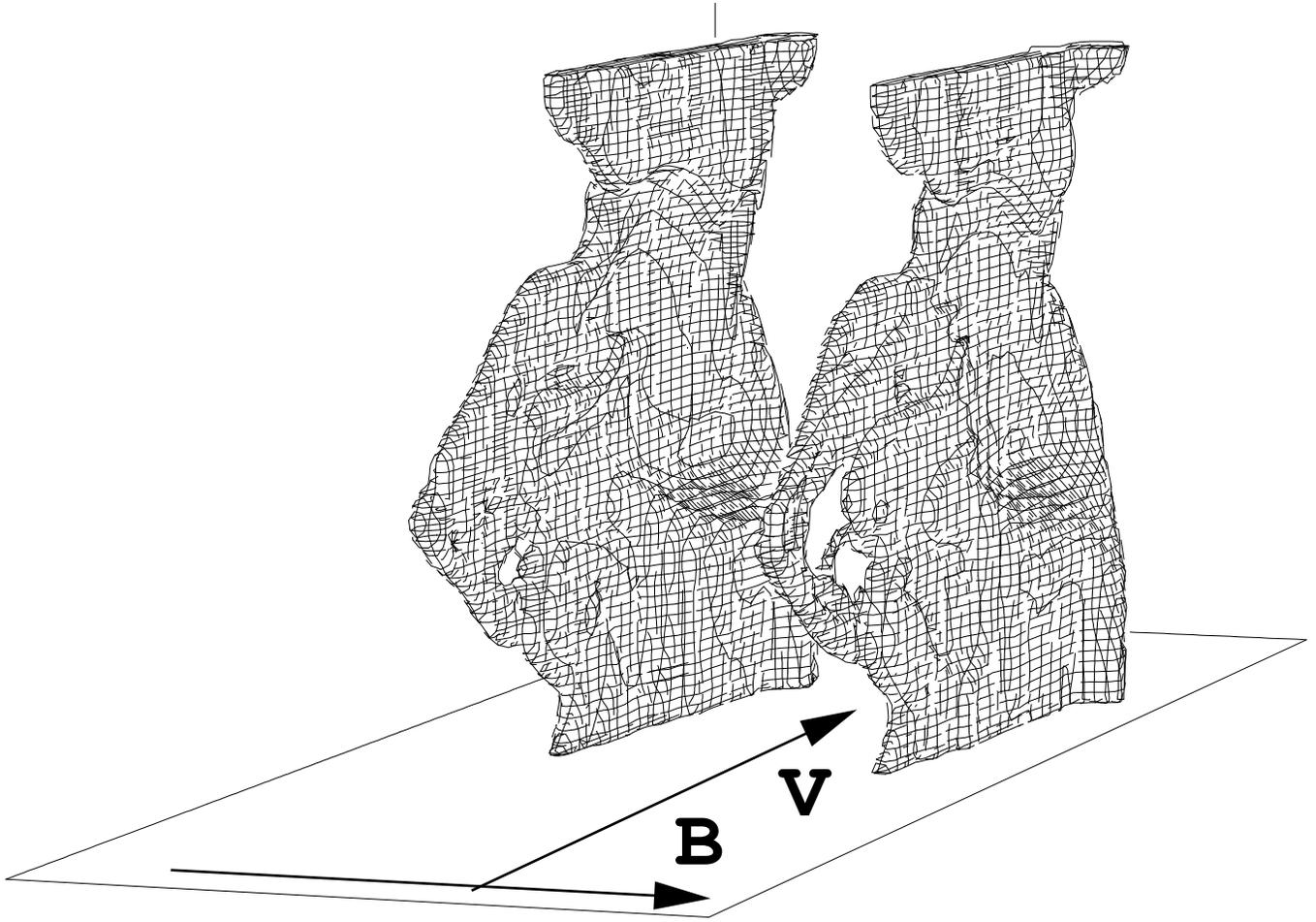
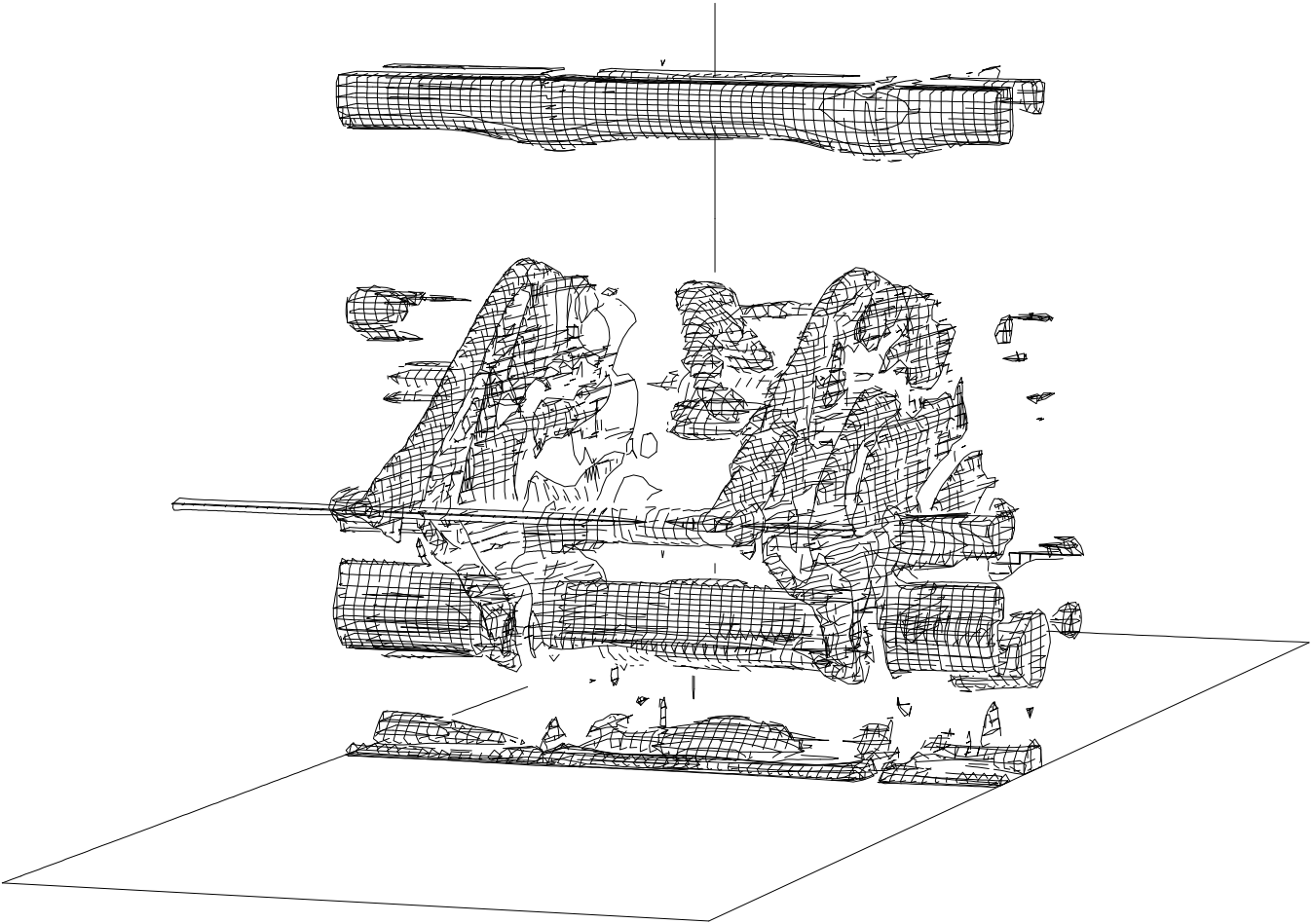
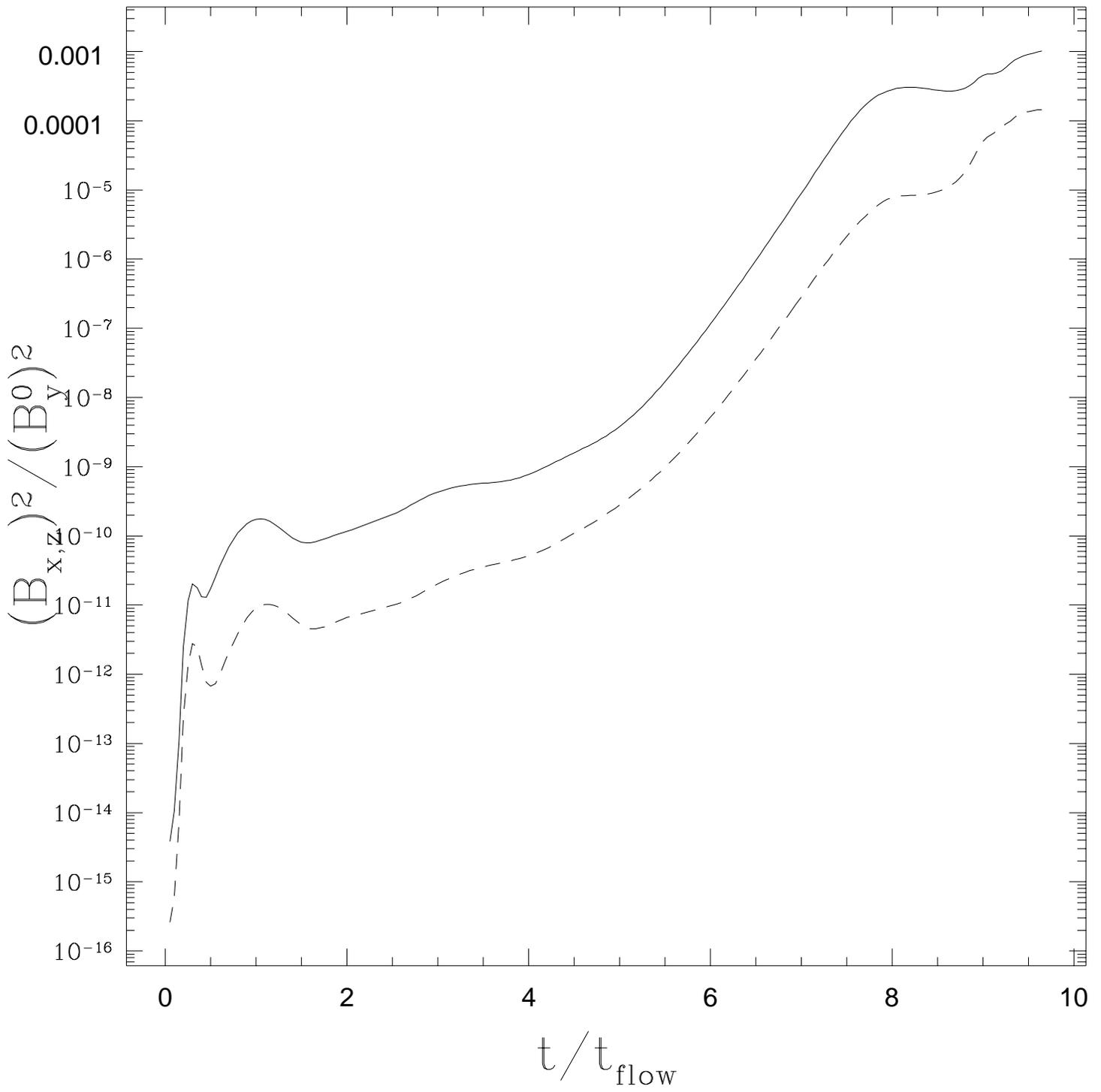
\end{document}